\documentclass{elsart}

\usepackage[square,comma]{natbib}
\usepackage{graphicx}
\usepackage{pxfonts}
\usepackage{lineno}

\usepackage{amssymb}
\usepackage{footnote}
\usepackage{multirow}
\usepackage{varwidth}
\usepackage{array}

\journal{}

\begin{document}

\thispagestyle{empty}
\begin{Large}
\textbf{DEUTSCHES ELEKTRONEN-SYNCHROTRON}

\textbf{\large{Ein Forschungszentrum der Helmholtz-Gemeinschaft}\\}
\end{Large}

DESY 13-101

June 2013

\begin{eqnarray}
\nonumber
\end{eqnarray}
\begin{center}
\begin{Large}
\textbf{Proposal for a scheme to generate 10 TW-level femtosecond
x-ray pulses for imaging single protein molecules at the European
XFEL}
\end{Large}
\begin{eqnarray}
\nonumber &&\cr \nonumber
\end{eqnarray}

\begin{large}
Svitozar Serkez$^a$, Vitali Kocharyan$^a$, Evgeni Saldin$^a$, Igor
Zagorodnov$^a$, Gianluca Geloni$^b$, and Oleksander Yefanov$^c$
\end{large}

\textsl{\\$^a$Deutsches Elektronen-Synchrotron DESY, Hamburg}
\begin{large}

\end{large}
\textsl{\\$^b$European XFEL GmbH, Hamburg}
\begin{large}

\end{large}
\textsl{\\$^c$Center for Free-Electron Laser Science, Hamburg}
\begin{eqnarray}
\nonumber
\end{eqnarray}
\begin{eqnarray}
\nonumber
\end{eqnarray}
ISSN 0418-9833
\begin{eqnarray}
\nonumber
\end{eqnarray}
\begin{large}
\textbf{NOTKESTRASSE 85 - 22607 HAMBURG}
\end{large}
\end{center}
\clearpage
\newpage
\begin{frontmatter}



\title{Proposal for a scheme to generate 10 TW-level femtosecond x-ray pulses for imaging single protein molecules at the European XFEL}


\author[DESY]{Svitozar Serkez \thanksref{corr},}
\thanks[corr]{Corresponding Author. E-mail address: svitozar.serkez@desy.de}
\author[DESY]{Vitali Kocharyan,}
\author[DESY]{Evgeni Saldin,}
\author[DESY]{Igor Zagorodnov,}
\author[XFEL]{Gianluca Geloni,}
\author[CFEL]{and Oleksander Yefanov}

\address[DESY]{Deutsches Elektronen-Synchrotron (DESY), Hamburg, Germany}
\address[XFEL]{European XFEL GmbH, Hamburg, Germany}
\address[CFEL]{Center for Free-Electron Laser Science, Hamburg, Germany}

\begin{abstract}
Single biomolecular imaging using XFEL radiation is an emerging
method for protein structure determination using the "diffraction
before destruction" method at near atomic resolution. Crucial
parameters for such bio-imaging experiments are photon energy range,
peak power, pulse duration, and transverse coherence. The largest
diffraction signals are achieved at the longest wavelength that
supports a given resolution, which should be better than $0.3$ nm.
We propose a configuration which combines self-seeding and undulator
tapering techniques with the emittance-spoiler method in order to
increase the XFEL output peak power and to shorten the pulse
duration  up to a level sufficient for performing bio-imaging of
single protein molecules at the optimal photon energy range, i.e.
around 4 keV. Experiments at the LCLS confirmed the feasibility of
these three new techniques. Based on start-to-end simulations we
demonstrate that self-seeding, combined with undulator tapering,
allows one to achieve up to a 100- fold increase in a peak-power.  A
slotted foil in the last bunch compressor is added for x-ray pulse
duration control. Simulations indicate that one can achieve
diffraction to the desired resolution with 50 mJ (corresponding to
$10^{14}$ photons) per 10 fs pulse at 3.5 keV photon energy in a 100
nm focus. This result is exemplified using the photosystem I
membrane protein as a case study.
\end{abstract}

%
%
%
\end{frontmatter}



\section{\label{sec:intro} Introduction}

Infrastructure of the European XFEL facility offer a unique
opportunity to build, potentially, a 10 TW-level x-ray source
optimized for single biomolecule imaging. Crucial parameters for
this application are photon energy range, peak power, pulse
duration, and transverse coherence \cite{HAJD}-\cite{BERG}. In fact,
experimental requirements imply very demanding characteristics for
the radiation pulse. In particular, the x-ray beam should be
delivered in 10 fs-long pulses in the 10 TW-level, and within a
photon energy range between 3 keV and 5 keV.

The baseline SASE undulator sources at the European XFEL will
saturate at about $50$ GW \cite{ETDR}. While this limit is very far
from the 10 TW-level required for imaging of single biomolecules,
there is a cost-effective way to improve the output power, when the
FEL undulators are longer than the saturation length. All the
requirements for single molecular imaging in terms of photon beam
characteristics can be satisfied by a simple combination of
self-seeding \cite{SELF}-\cite{ASYM}, emittance spoiler foil
\cite{EMM1}-\cite{DING}, and undulator tapering techniques
\cite{TAP1}-\cite{LAST}. Relying on these techniques we discuss a
scheme of operation for a bio-imaging undulator source, which could
be built at the European XFEL  based on start-to-end-simulations for
an electron beam with 1 nC charge \cite{S2ER}.  We demonstrate  that
it is possible to achieve up to a 100-fold increase in peak power of
the x-ray pulses: the x-ray beam would be delivered in about 10
fs-long pulses with 50 mJ energy each at photon energies around 4
keV.

\section{\label{sec:setup} Setup description}

\begin{figure}
\begin{center}
\includegraphics[clip, width=0.75\textwidth]{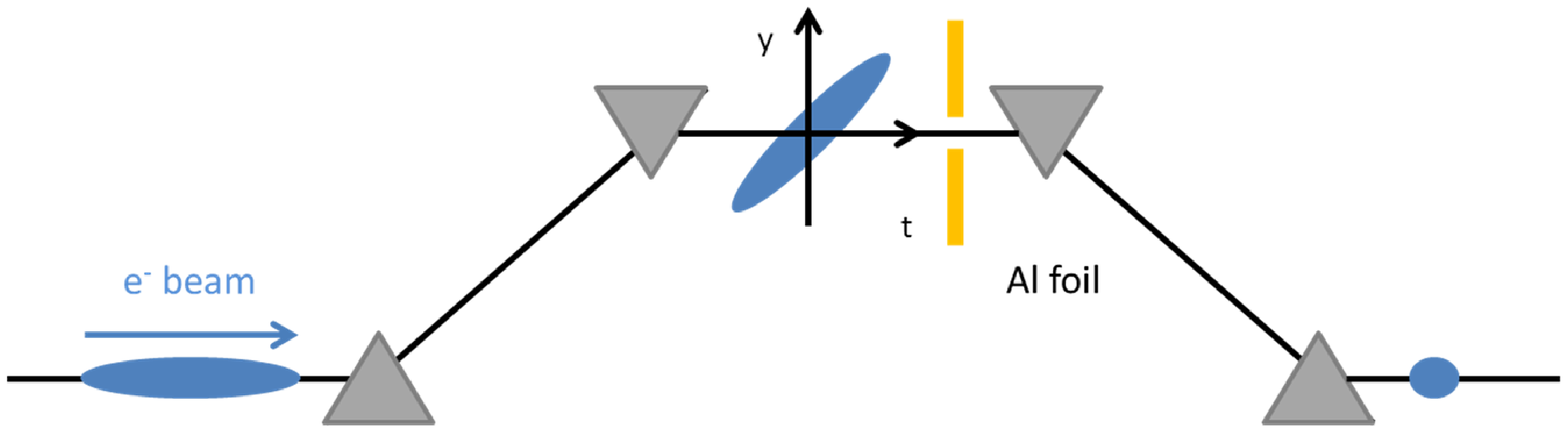}
\end{center}
\caption{Sketch of an electron bunch at the center of  the magnetic
bunch compressor chicane (adapted from \cite{EMM2}).} \label{slot1}
\end{figure}

In order to provide bio-imaging capabilities, x-ray pulses should be
provided with a tunable duration between 3 fs and 30 fs. While
proposals exist to tune photon pulses at the European XFEL in this
range, they require installation of additional hardware in the
undulator system \cite{BIO1,BIO3}. Here we exploit a simpler method
to reach the same results still assuming that the undulator system
is long enough\footnote{40 undulator cells}, but making only minimal
changes in the undulator system\footnote{Only a single-chicane
self-seeding setup with crystal monochromator is needed}. A proposal
\cite{EMM1,EMM2} and an experimental verification \cite{DING} have
been made in order to generate femtosecond x-ray pulses at the LCLS
by using a slotted spoiler foil located in the center of the last
bunch compressor. The method takes advantages of the high
sensitivity of the FEL gain process to the transverse emittance of
the electron bunch. By spoiling the emittance of most of the beam
while leaving a short unspoiled temporal slice, one can produce an
x-ray FEL pulse much shorter than in the case when the original
electron bunch is sent through the undulator.

\begin{figure}
\begin{center}
\includegraphics[ width=0.75\textwidth]{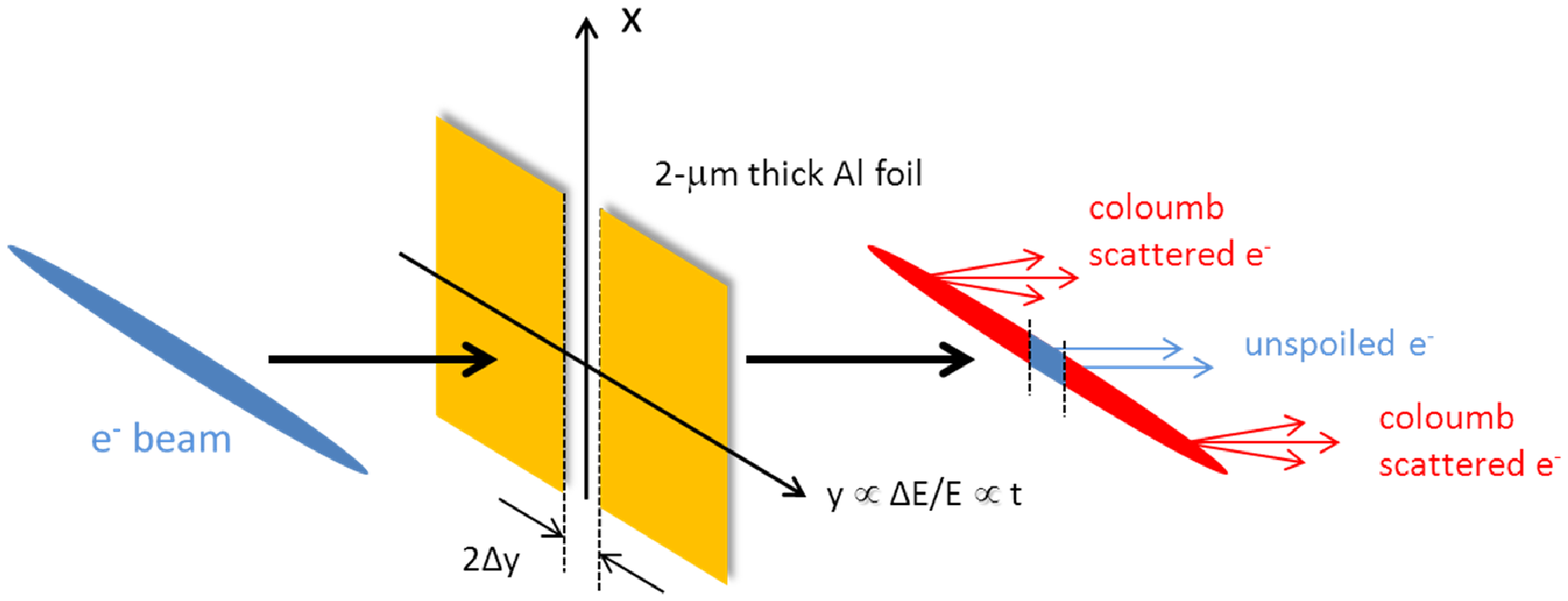}
\end{center}
\caption{ The slotted foil at chicane center generates a narrow,
unspoiled beam center (adapted from \cite{EMM2})} \label{slot2}
\end{figure}

Fig. \ref{slot1} shows a sketch of the slotted  foil  at the center
of the third and last bunch compressor BC3 at the European XFEL. The
last linac section before the third bunch compressor BC3 is set at
an off-crest accelerating rf phase, so that the beam energy at the
entrance of BC3 is correlated with time. Due to chromatic
dispersion, this chirp transforms into in a $y-t$ bunch tilt in the
chicane. At the center of BC3, i.e. at the point of maximum tilt, a
thin foil is placed in the path of the beam. The foil has a narrow
slot at its center, oriented as shown in Fig. \ref{slot2}. Coulomb
scattering of the electrons passing through the foil increases the
horizontal and vertical emittances of most of the beam, but leaves a
thin unspoiled slice where the beam passes through the slit, Fig.
\ref{slot2}.

\begin{figure}
\includegraphics[width=0.50\textwidth]{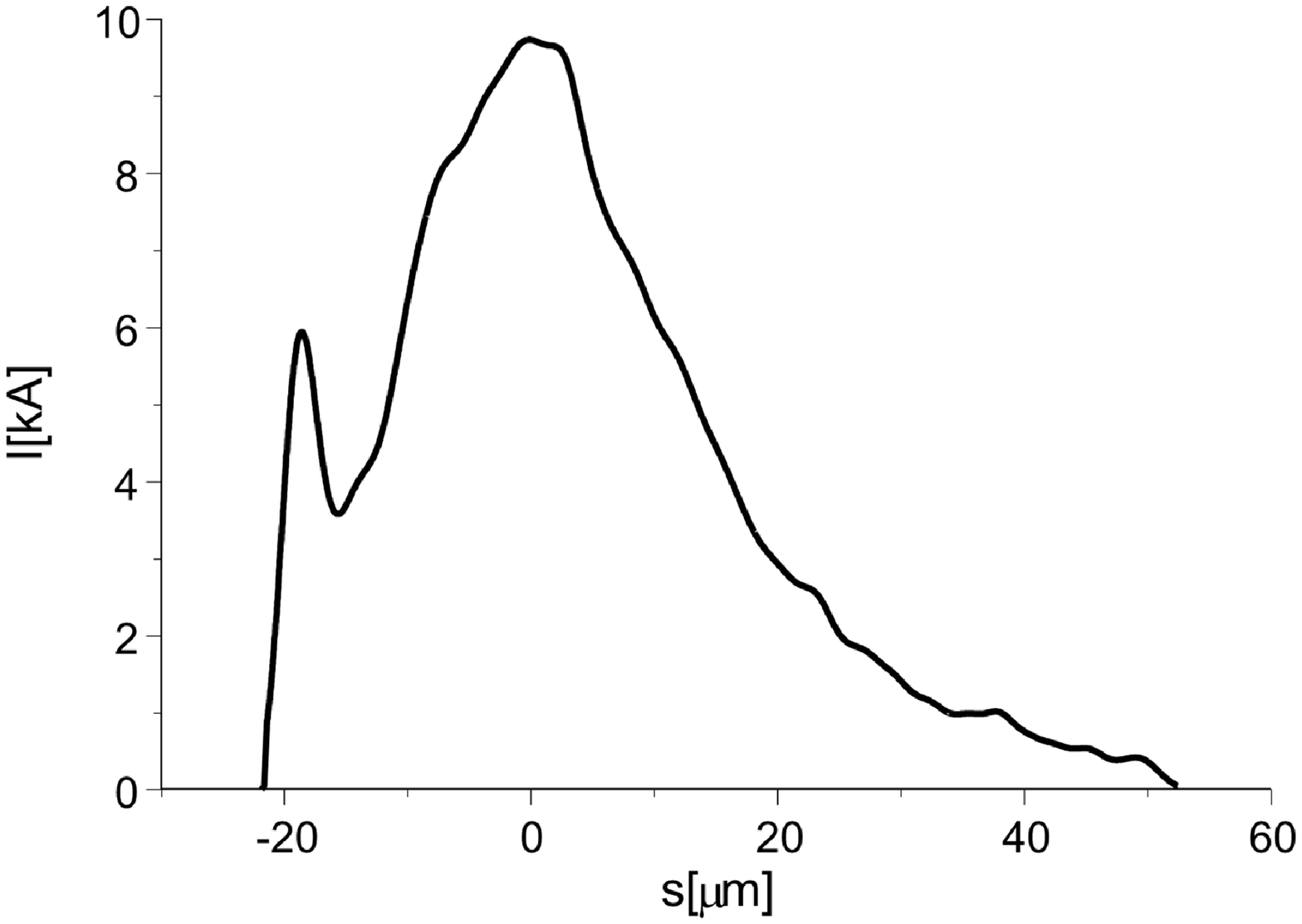}
\includegraphics[clip, width=0.50\textwidth]{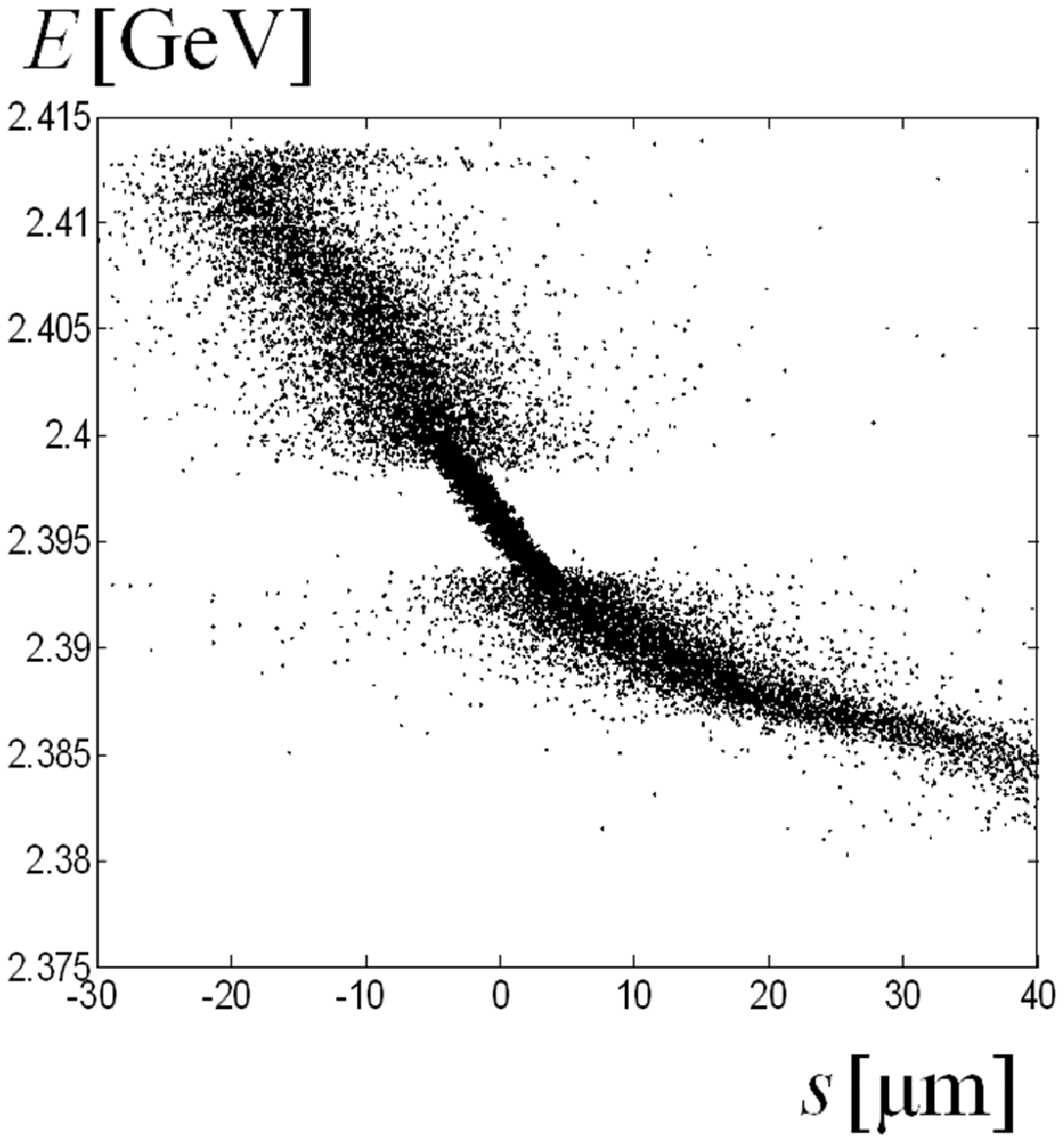}
\caption{Left plot: Current profile after BC3 without foil. Right
plot: Longitudinal phase space distribution of the particles after
BC3, with foil. The simulation includes multiple Coulomb scattering
in a $2\mu$m thin aluminum foil with a slot width of $0.7$ mm.}
\label{currsl}
\end{figure}

\begin{figure}
\includegraphics[width=0.50\textwidth]{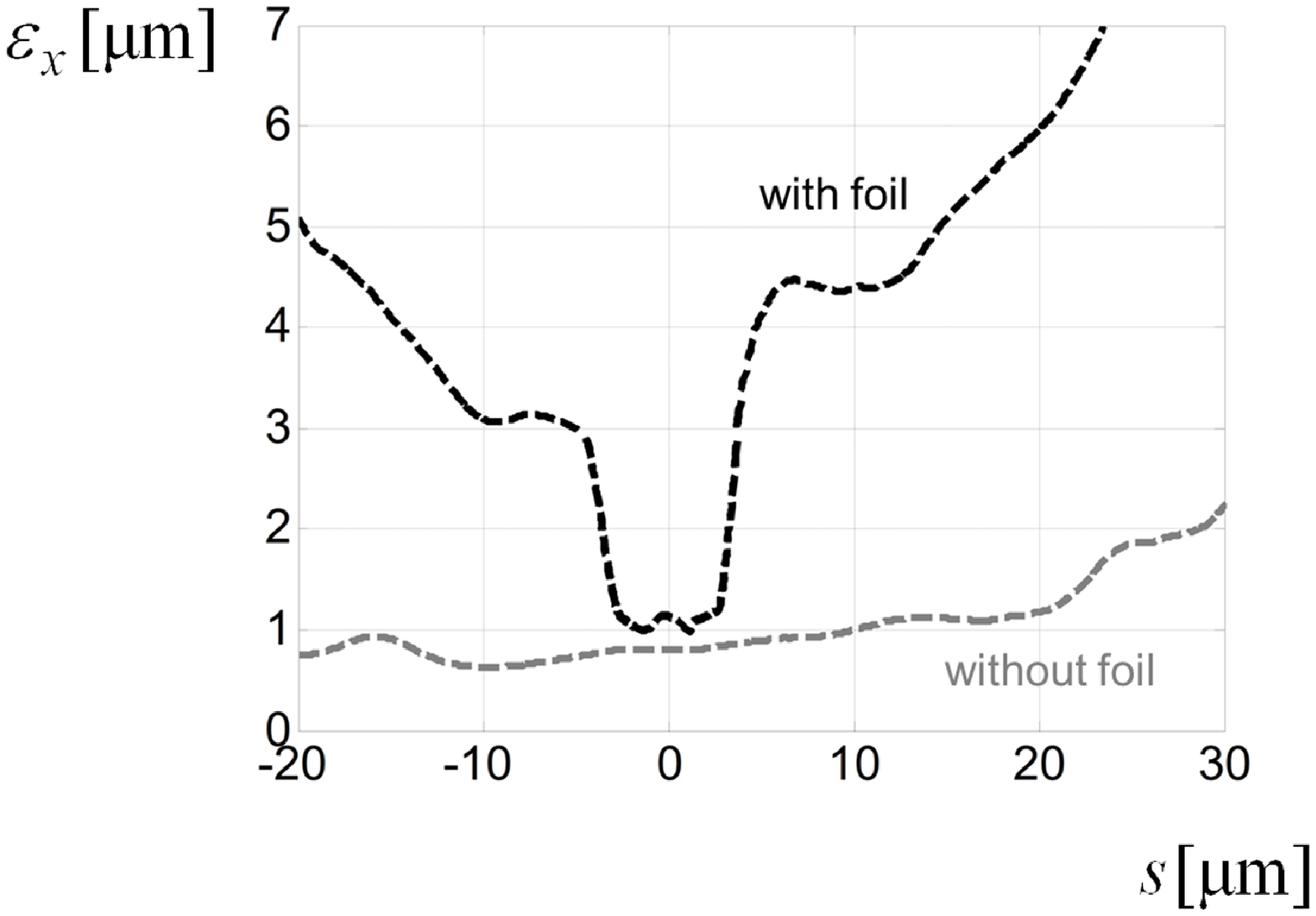}
\includegraphics[width=0.50\textwidth]{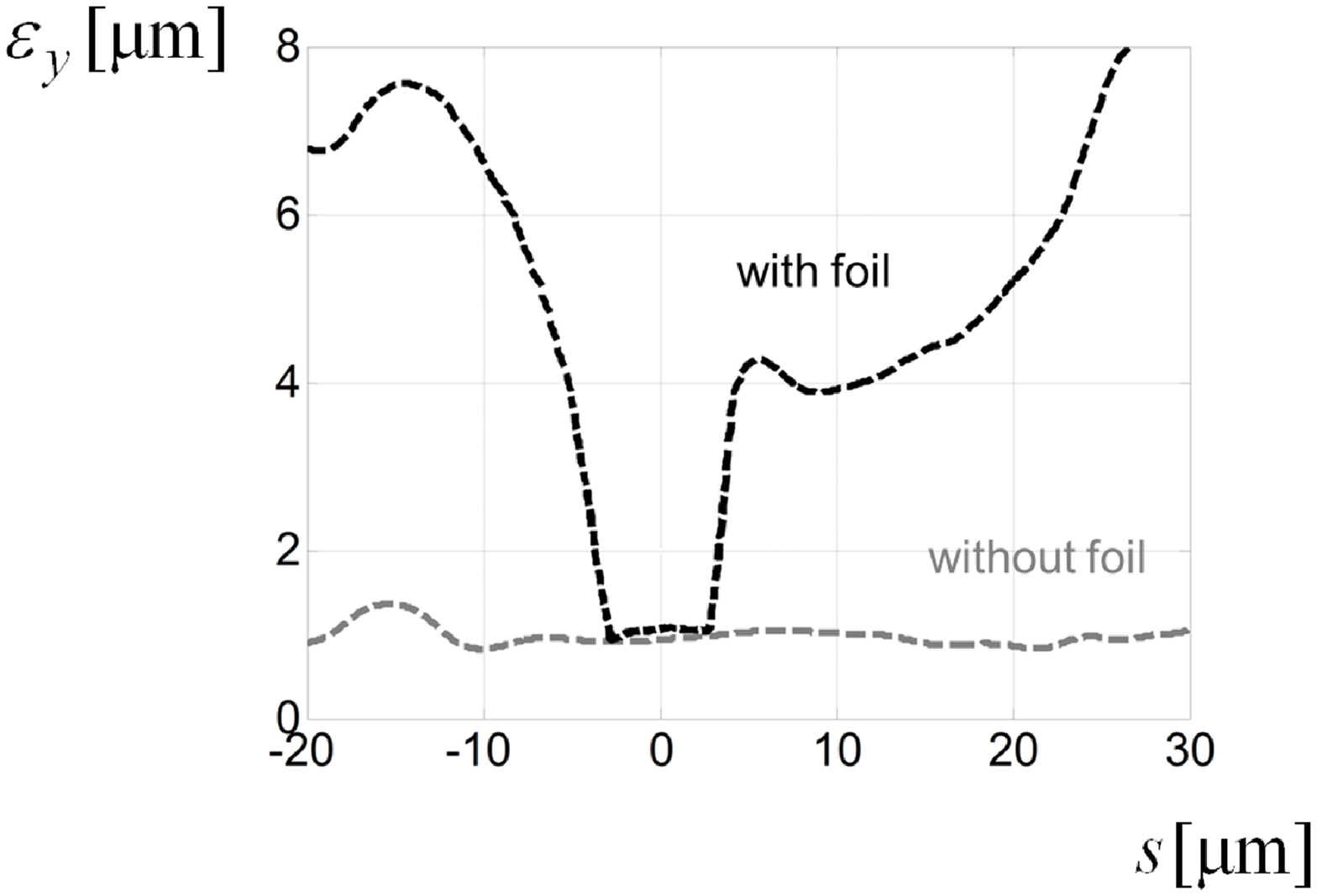}
\caption{Left plot: Vertical normalized emittance as a function of
the position inside the electron bunch after BC3. The grey dashed
curve is from particle tracking without foil. The black dashed curve
is from particle tracking with foil. Right plot: Horizontal
emittance as a function of the position inside the electron bunch
after BC3. The grey dashed curve is from particle tracking without
foil. The black dashed curve is from particle tracking with foil.
(In both plots we removed 6 $\%$ of strongly scattered particles
from the analysis.) } \label{emxy}
\end{figure}

Spoiling the emittance of most of the beam by a factor $\sim 5$
strongly suppresses the FEL gain, while the short, unspoiled
temporal slice produces an x-ray FEL pulse much shorter than the
FWHM electron bunch duration, Fig. \ref{currsl} (left plot). If a
very narrow slit is used, uncorrelated energy spread and betatron
beam size dominate the output slice length \cite{EMM1,EMM2}, and one
obtains a nonlinear growth of the x-ray pulse length versus the slot
width. When the slit becomes larger, the growth becomes linear and
the x-ray pulse length is mainly determined by the width of the slit
\cite{DING}. The minimum duration of the unspoiled slice of the
electron bunch measured at the LCLS is about 3 fs \cite{DING}. We
use current profile, normalized emittance, energy spread profile,
electron beam energy spread and wakefields from \cite{S2ER}. The
electron beam charge is 1 nC, and the peak current is 10 kA, Fig.
\ref{currsl} (left plot). Detailed computer simulations with $2\cdot
10^5$ macroparticles have been carried out to evaluate the
performance of the slotted spoiler using the tracking code ELEGANT
\cite{ELEG}. They include multiple Coulomb scattering in a $2\mu$m
thin aluminum foil. The longitudinal distribution of the particles
just after the BC3 chicane is shown in Fig \ref{currsl} (right
plot). A slit full-width of $0.7$ mm selects a small fraction of
electrons, about $20 \%$, and produces an unspoiled electron bunch
slice after BC3, with a duration of about $18$ fs FWHM\footnote{As
we will see, the FEL gain-narrowing allows an x-ray pulse duration
of about $12$ fs}, Fig. \ref{emxy}.

\begin{figure}
\begin{center}
\includegraphics[clip, width=0.75\textwidth]{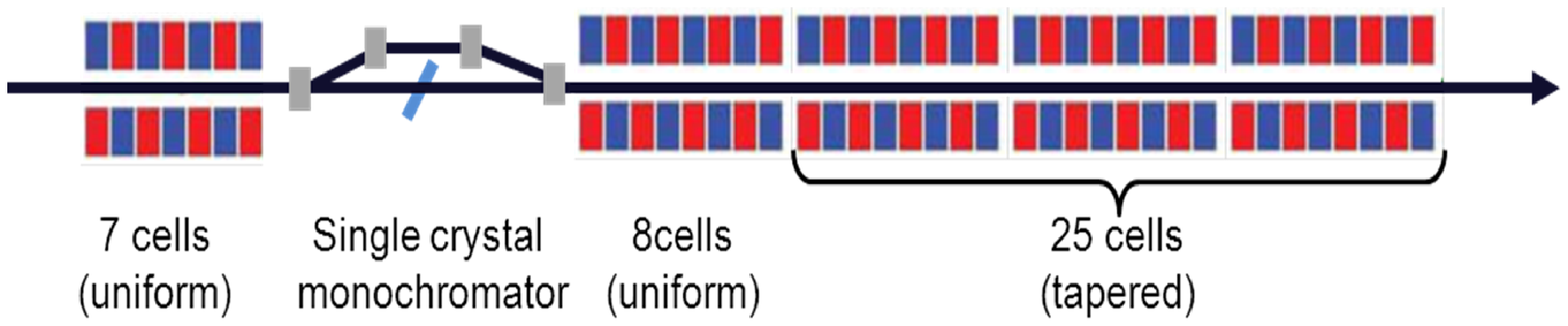}
\end{center}
\caption{Scheme for a 10 TW-power level undulator source.
Self-seeding and undulator tapering greatly improve the FEL
efficiency. X-ray pulse length control is obtained using a slotted
foil in the last bunch compressor. The magnetic chicane accomplishes
three tasks by itself.  It creates an offset for single crystal
monochromator, it removes the electron microbunching produced in the
upstream undulator, and it acts as a magnetic delay line.}
\label{layout}
\end{figure}
A design of a self-seeding setup based on the undulator system for
the European XFEL is sketched in Fig. \ref{layout}. The method for
generating high power x-ray pulses exploits a combination of a
self-seeding scheme with an undulator tapering technique. Tapering
consists in a slow reduction of the field strength of the undulator
in order to preserve the resonance wavelength, while the kinetic
energy of the electrons decreases due to the FEL process. The
undulator taper could be simply implemented at discrete steps from
one undulator segment to the next, by changing the undulator gap.
Highly monochromatic pulses generated with the self-seeding
technique make the tapering much more efficient than in the SASE
case.

Here we study a scheme for generating 10 TW-level x-ray pulses in a
SASE3-type tapered undulator. However, a similar scheme can be
implemented at SASE1 or SASE2.  In the following we assume to have
$40$ undulator segments at our disposal.  We optimize our setup
based on start-to-end simulations for a 17.5 GeV electron beam with
1 nC charge compressed up to 10 kA peak current. In this way, the
output power of the SASE3 undulator could be increased from the
value of 100 GW in the SASE regime\footnote{There is an increase in
the SASE saturation power with respect to the nominal mode of
operation, due to the increase in peak current from the nominal 5 kA
to our case, where 10 kA are obtained with particular settings of
the bunch formation system.} to about 5 TW in the photon energy
range around 4 keV.

\begin{figure}
\begin{center}
\includegraphics[clip, width=0.75\textwidth]{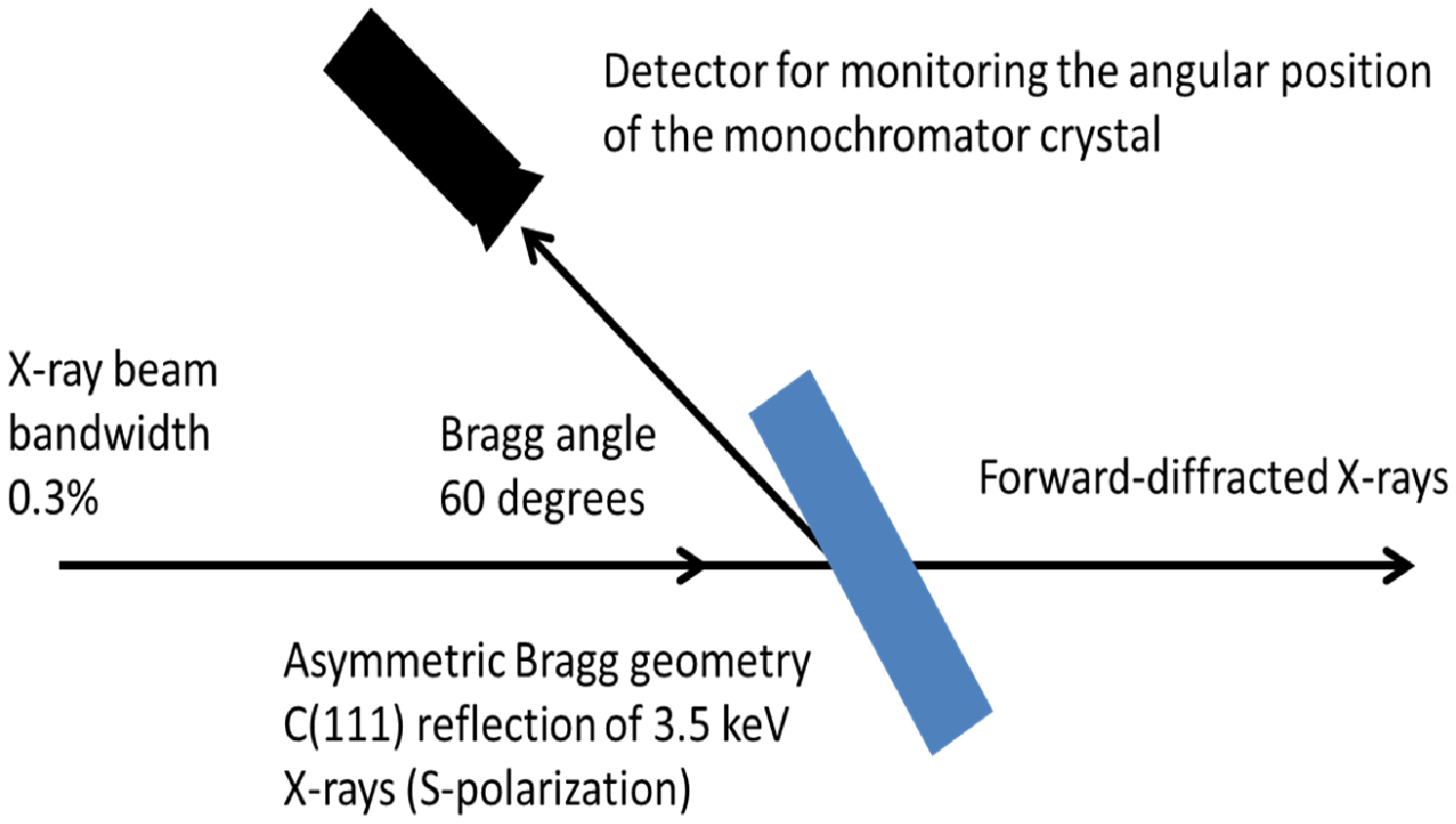}
\end{center}
\caption{Schematic of the single crystal monochromator for operation
in the photon energy range around 4 keV.} \label{cryg}
\end{figure}
\begin{figure}
\begin{center}
\includegraphics[clip, width=0.30\textwidth]{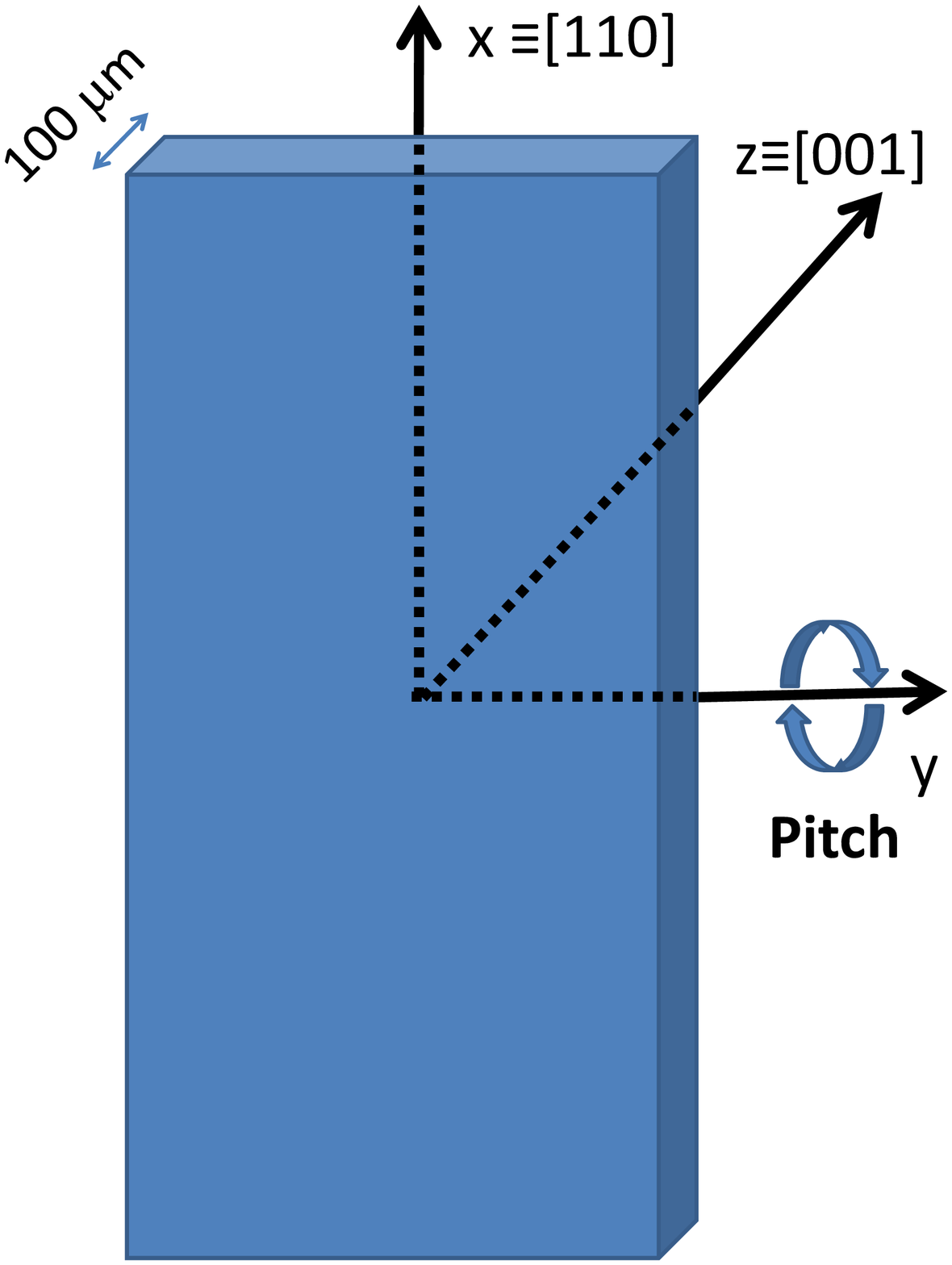}
\end{center}
\caption{Drawing of the orientation of the diamond crystal
considered in the article.} \label{cryaxes}
\end{figure}
\begin{figure}
\begin{center}
\includegraphics[clip, width=0.5\textwidth]{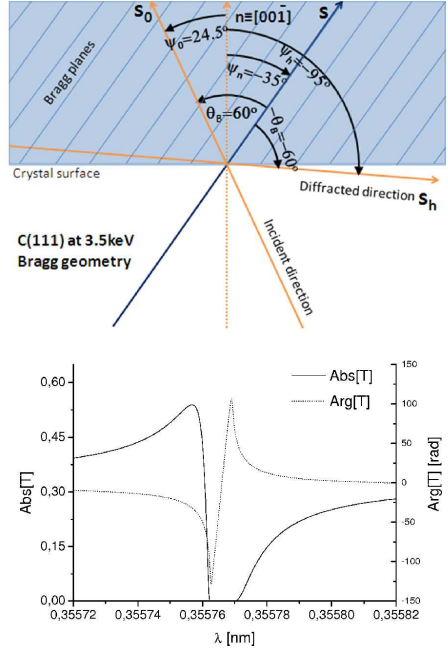}
\end{center}
\caption{Upper plot: scattering geometry (we are following the
notation in \cite{AUTH}). Lower plot: modulus and phase of the
transmittance for the C(111) asymmetric Bragg reflection  from the
diamond crystal in Fig. \ref{cryaxes} at 3.5 keV.} \label{C111}
\end{figure}
Our  design adopts the simplest self-seeding scheme, which uses the
transmitted x-ray beam from a single crystal to seed the same
electron bunch, Fig. \ref{cryg}. In the following we will consider a
$100~\mu$m-thick diamond crystal. We  define a Cartesian reference
system [$x, y, z$] linked with the crystal. The direction $z$
corresponds to the direction identified by the Miller indexes [0, 0,
1], while $x$ and $y$ are specified as in Fig. \ref{cryaxes}.  The
crystal can rotate freely around the $y$ axis (pitch angle) as
indicated in figure. In this way we can exploit several symmetric
and asymmetric reflections. By changing the pitch angle of the
crystal in Fig. \ref{cryaxes} we are able, in fact, to cover the
entire energy range between 3 keV and 13 keV \cite{SHVI},
\cite{ASYM}. In the low energy range between 3 keV and 5 keV we use
a C(111) asymmetric reflection (in Bragg and Laue geometry,
depending on the energy). For self-seeding implementation, we are
interested in the forward diffracted beam. From this viewpoint, the
crystal can be characterized as a filter with given complex
transmissivity. In Fig. \ref{C111} we show scattering geometry,
amplitude and phase of the transmittance for the C(111) asymmetric
Bragg reflection at 3.5 keV.

Summing up, the overall self-seeding setup proposed here consists of
three parts: a SASE undulator, a self-seeding single crystal
monochromator and an output undulator, in which the monochromatic
seed signal is amplified up to the 10 TW-level, Fig. \ref{layout}.
Calculations show that, in order not to spoil the electron beam
quality and to simultaneously reach signal dominance over shot
noise, the number of cells in the first (SASE) undulator should be
equal to 7. The output undulator consists of two sections. The first
section is composed by an uniform undulator, the second section by a
tapered undulator. The monochromatic seed signal is exponentially
amplified passing through the first uniform part of the output
undulator. This section is long enough, 8 cells, in order to reach
saturation, which yields about 100 GW power. Finally, in the second
part of the output undulator the monochromatic FEL output is
enhanced up to 5 TW by taking advantage of the undulator magnetic
field taper over the last 25 cells.

\begin{figure}
\includegraphics[width=0.50\textwidth]{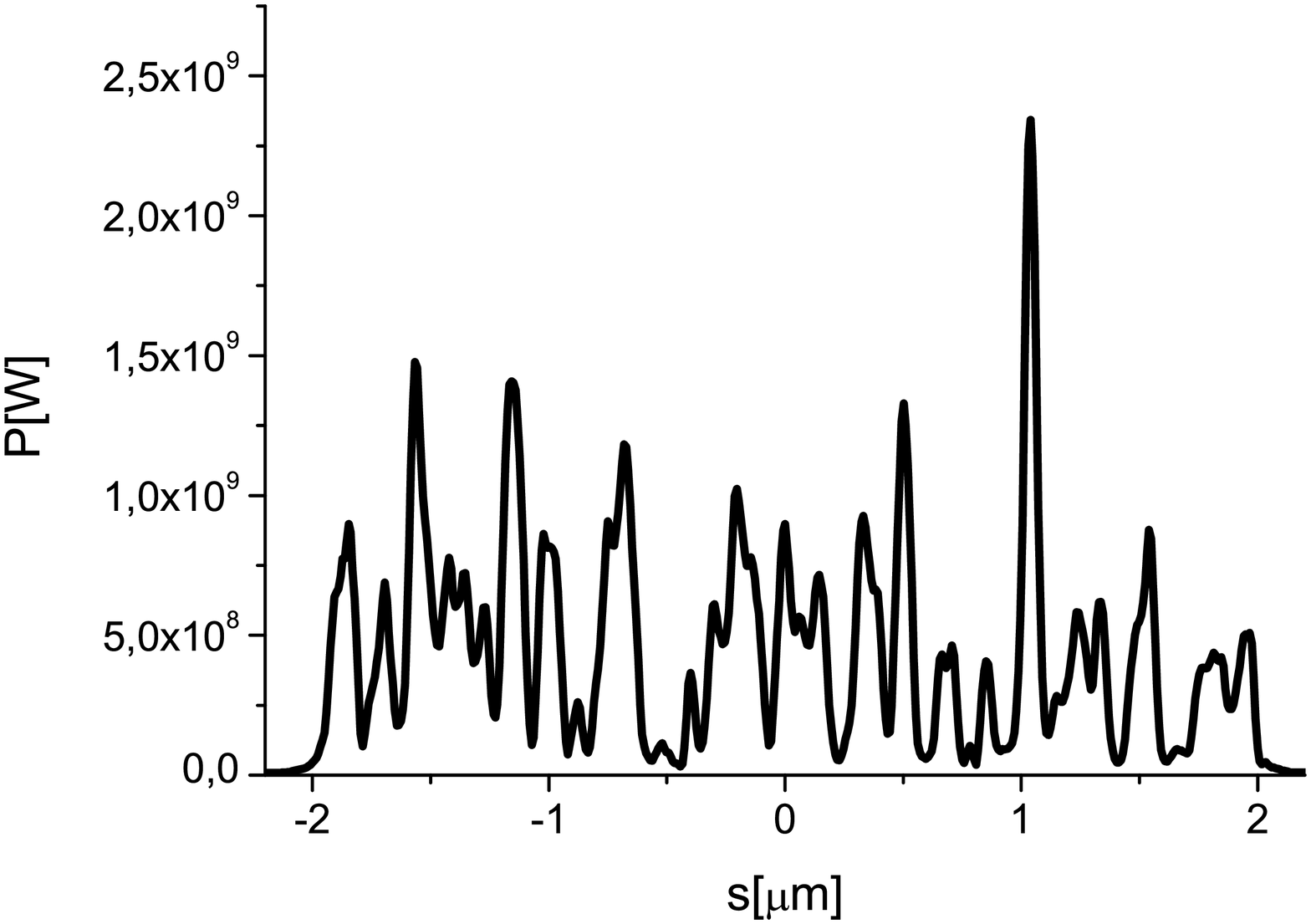}
\includegraphics[width=0.50\textwidth]{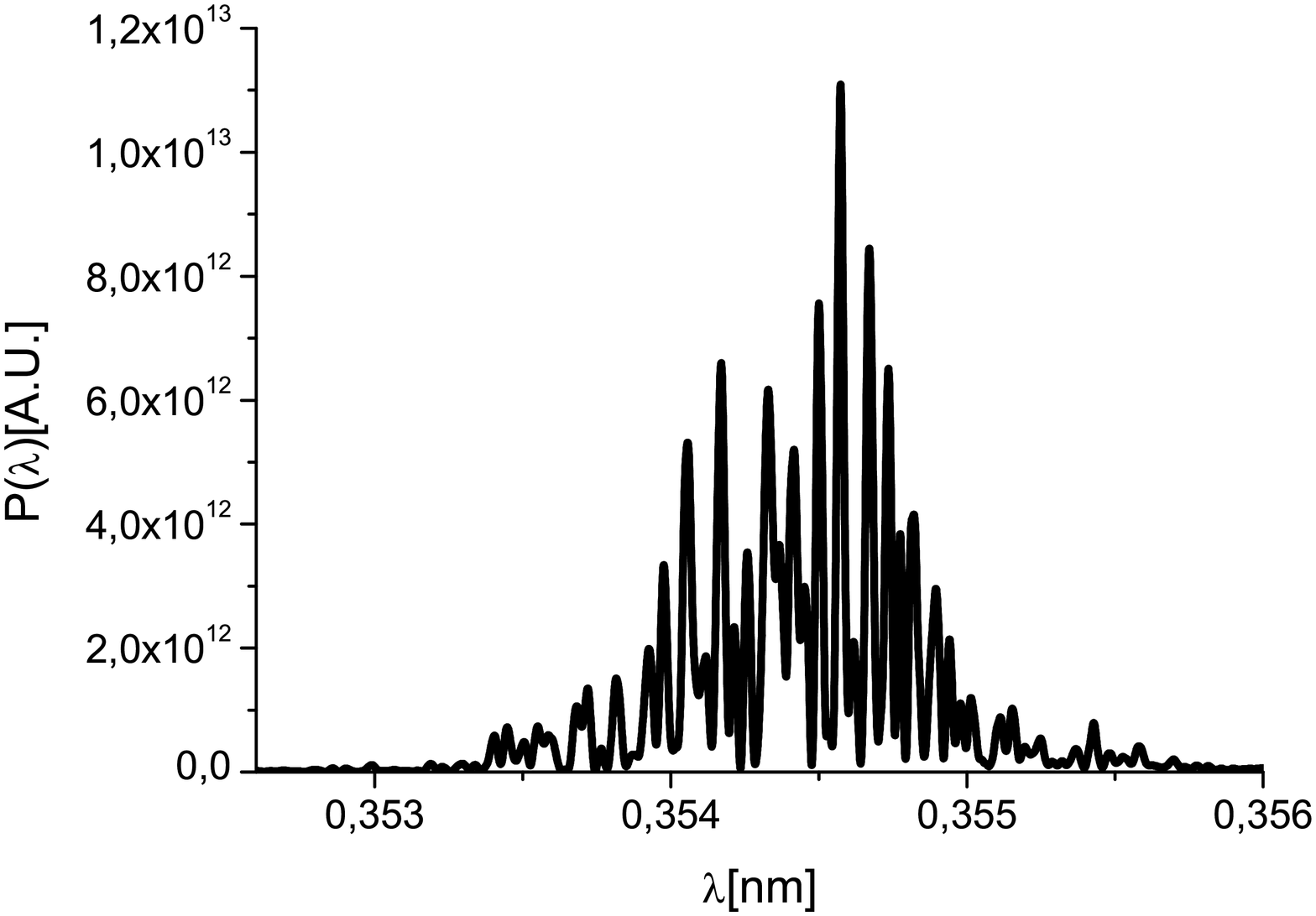}
\caption{Power distribution and spectrum of the SASE x-ray pulse at
the exit of the first undulator.} \label{PSpin1}
\end{figure}

\begin{figure}
\includegraphics[width=0.50\textwidth]{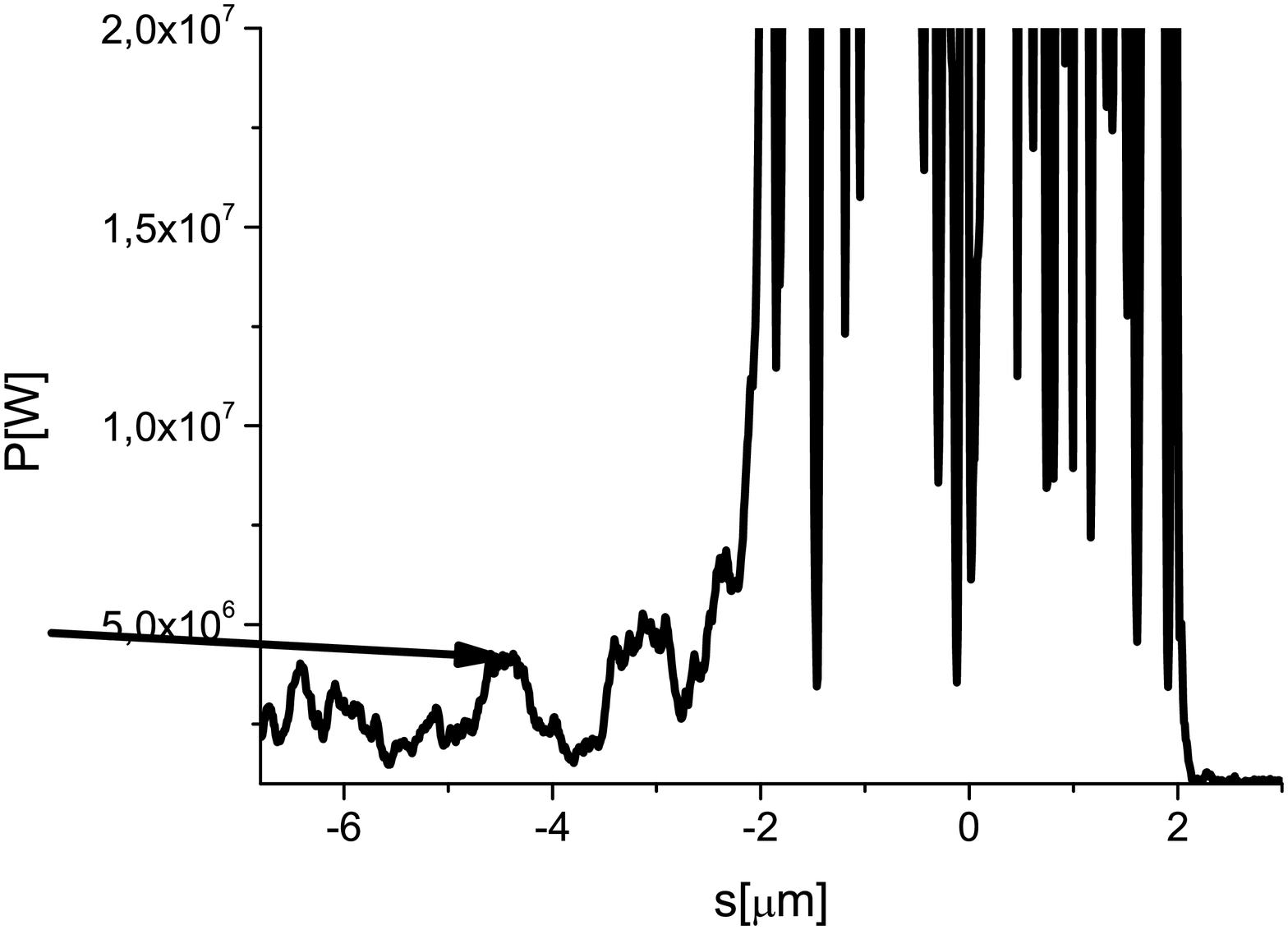}
\includegraphics[width=0.50\textwidth]{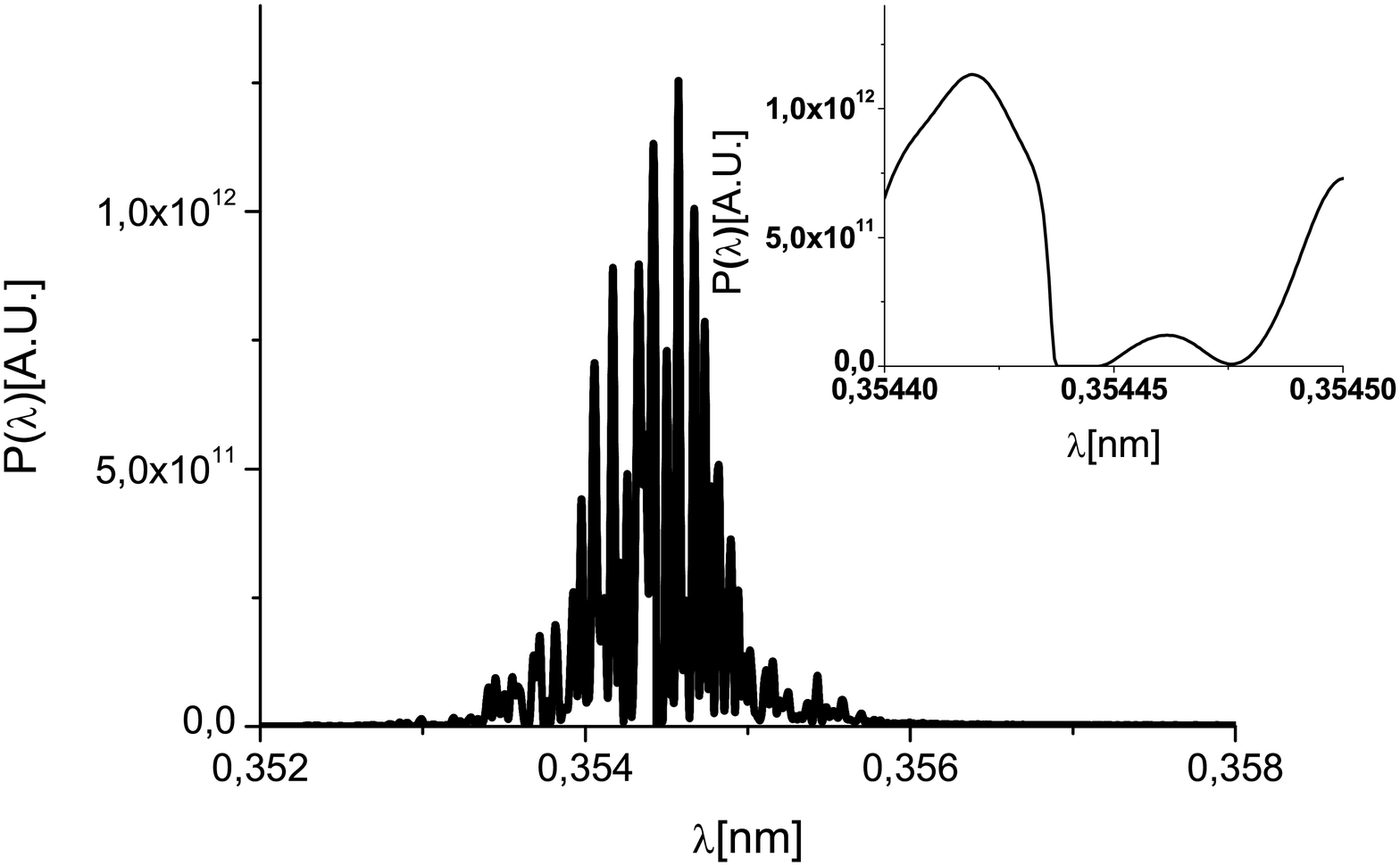}
\caption{Power distribution and spectrum of the SASE x-ray pulse at
after the wake monochromator. The seed pulse is indicated by an
arrow in the left plot.} \label{PSpseed1}
\end{figure}

\begin{figure}
\begin{center}
\includegraphics[width=0.50\textwidth]{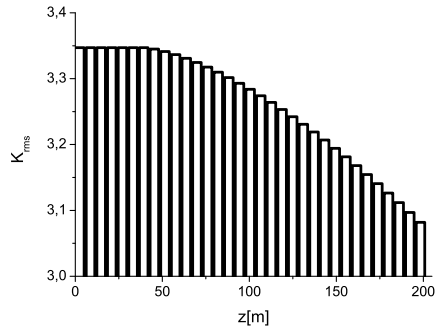}
\end{center}
\caption{Taper configuration for high-power mode of operation at
$0.35$ nm.} \label{Taplaw}
\end{figure}

\begin{figure}
\includegraphics[width=0.50\textwidth]{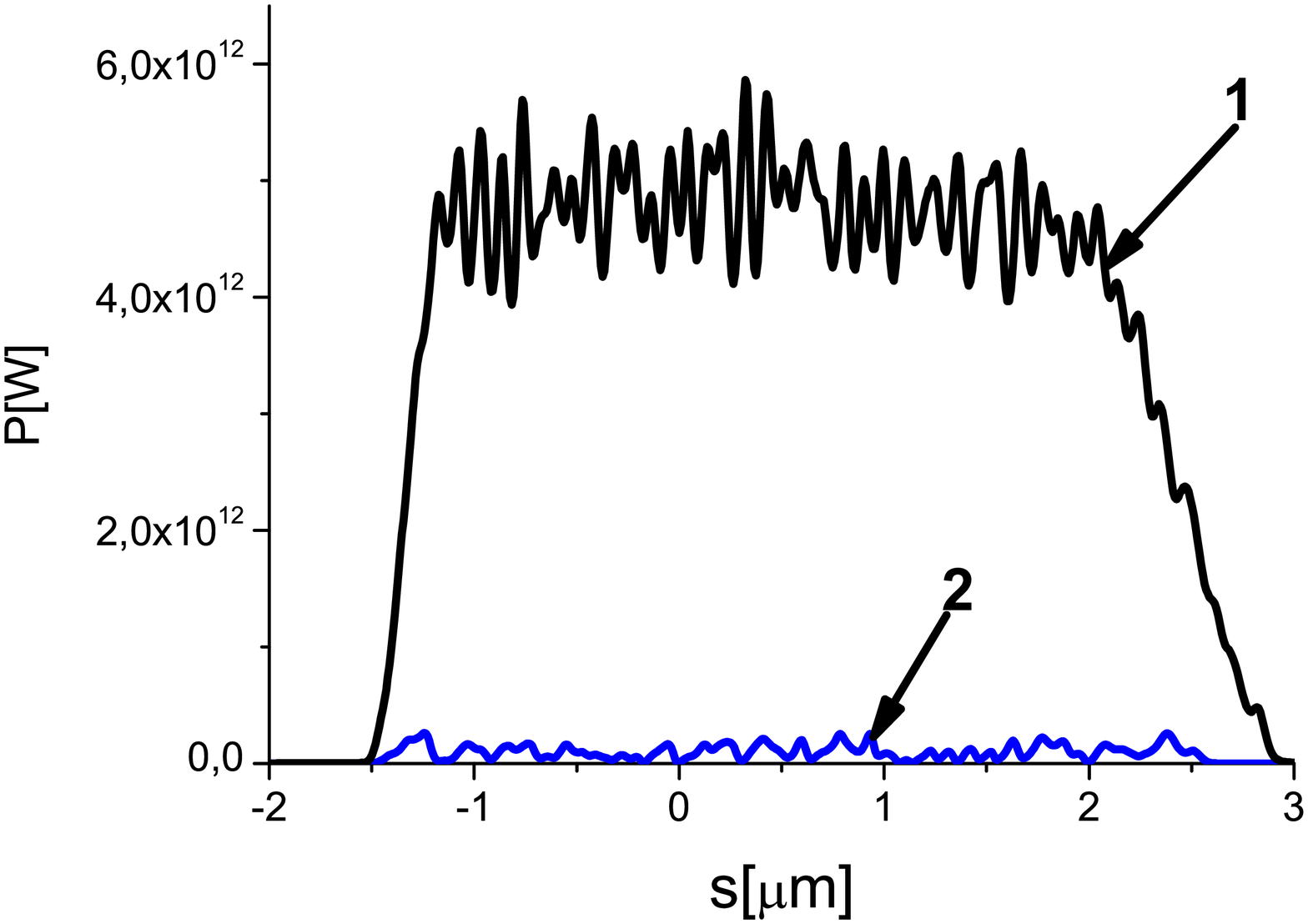}
\includegraphics[width=0.50\textwidth]{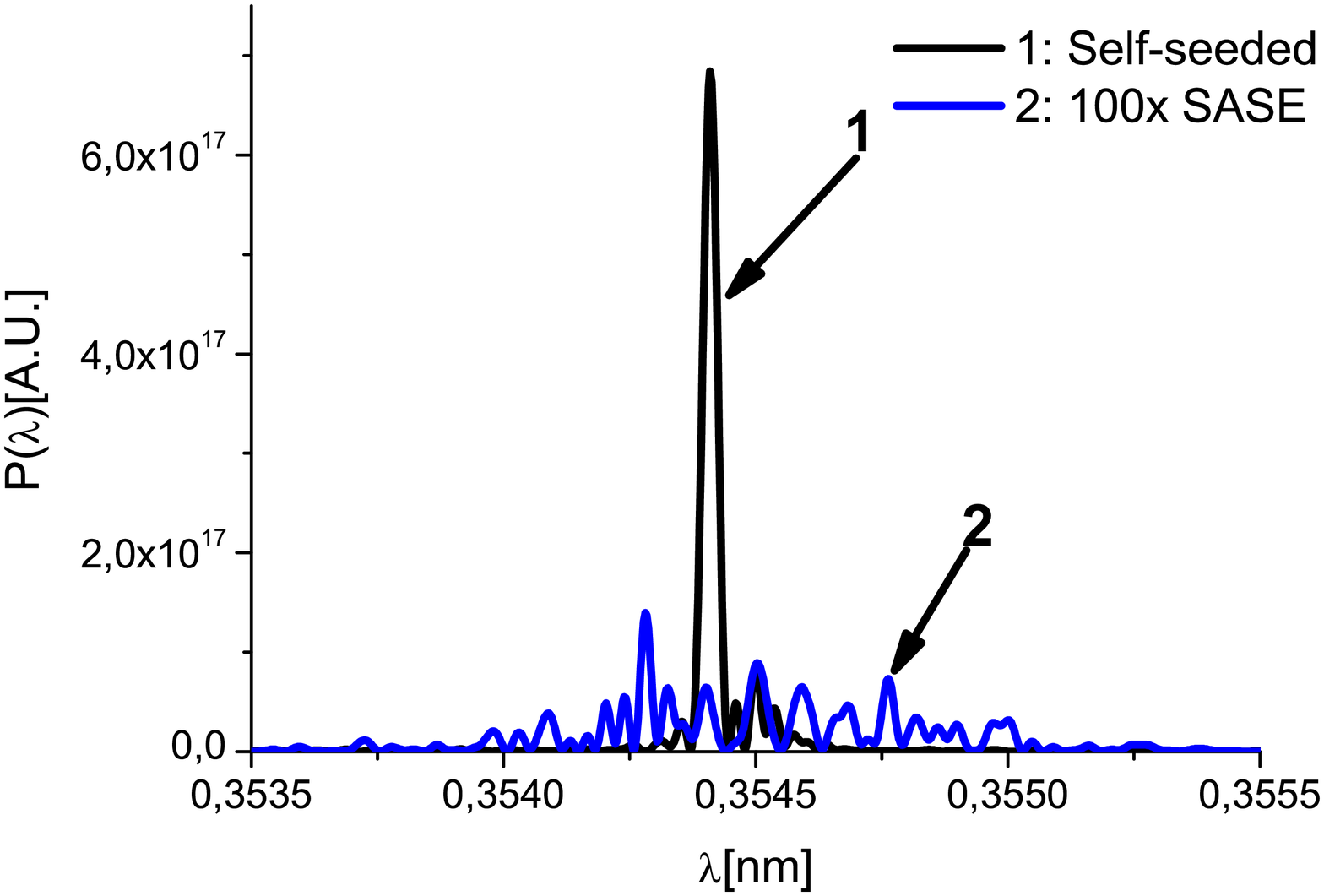}
\caption{Power distribution and spectrum of the output radiation
pulse. The self-seeded line 1 is compared with the SASE line 2,
showing the advantages of our method. The SASE spectrum is magnified
of a factor 100, to make it visible in comparison with the
self-seeded spectrum.} \label{PSpout1}
\end{figure}

\begin{figure}
\begin{center}
\includegraphics[width=0.50\textwidth]{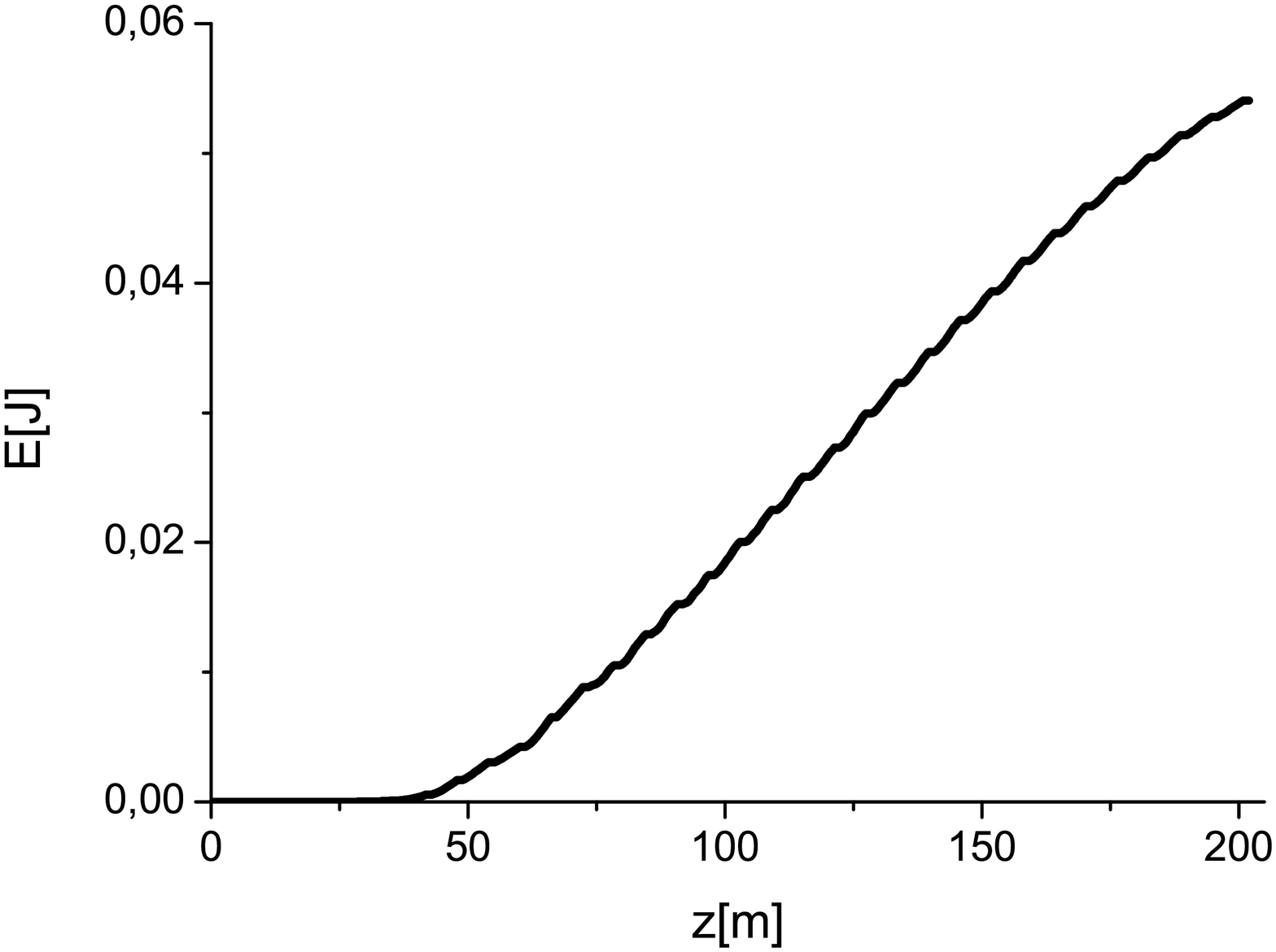}
\end{center}
\caption{Energy of the seeded FEL pulse as a function of the
distance inside the output undulator.} \label{Eout1}
\end{figure}

Simulations were performed with the help of the Genesis code
\cite{GENE} in the following way: first we calculated the 3D field
distribution at the exit of the first undulator, and dumped the
field file. Subsequently, we performed a temporal Fourier transform
followed by filtering through the monochromator, by using the filter
amplitude transfer function. The electron beam microbunching is
washed out by the presence of a nonzero momentum compaction factor
$R_{56}$ in the chicane. Therefore, for the second undulator we used
a beam file with no initial microbunching, but with characteristics
(mainly the energy spread) induced by the FEL amplification process
in the first SASE undulator. The amplification process in the second
undulator starts from the seed field file. Shot-noise initial
condition were included. The output power and spectrum after the
first SASE undulator tuned at 3.5 keV is shown in Fig. \ref{PSpin1}.
The crystal acts as bandstop filter, and the output spectrum is
plotted in Fig. \ref{PSpseed1} (right). The signal in the time
domain exhibits a long monochromatic tail, which is used for
seeding, Fig. \ref{PSpseed1} (left). The electron bunch is slightly
delayed by proper tuning of the self-seeding magnetic chicane, in
order to superimpose the unspoiled part of the electron bunch with
the seed signal. After saturation the undulator is tapered, i.e. the
undulator $K$ parameter is changed section by section, following the
configuration in Fig. \ref{Taplaw}.

The output power and spectrum of the entire setup, that is after the
second part of the output undulator is shown in Fig \ref{PSpout1}.
In particular, the self seeded power is compared with the SASE
power. The evolution of the output energy in the photon pulse is
plotted in Fig. \ref{Eout1}.

\begin{figure}[tb]
\includegraphics[width=0.5\textwidth]{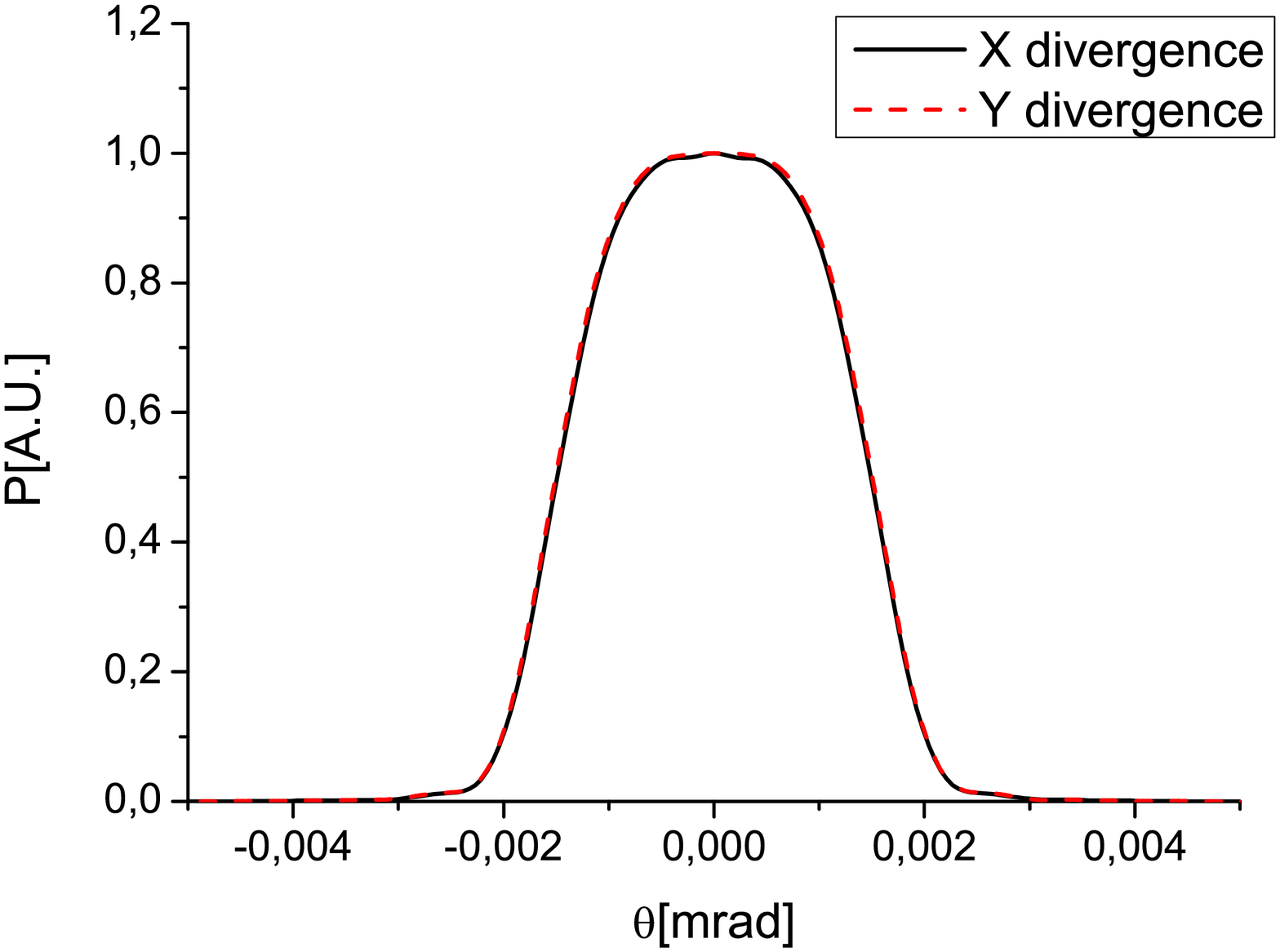}
\includegraphics[width=0.5\textwidth]{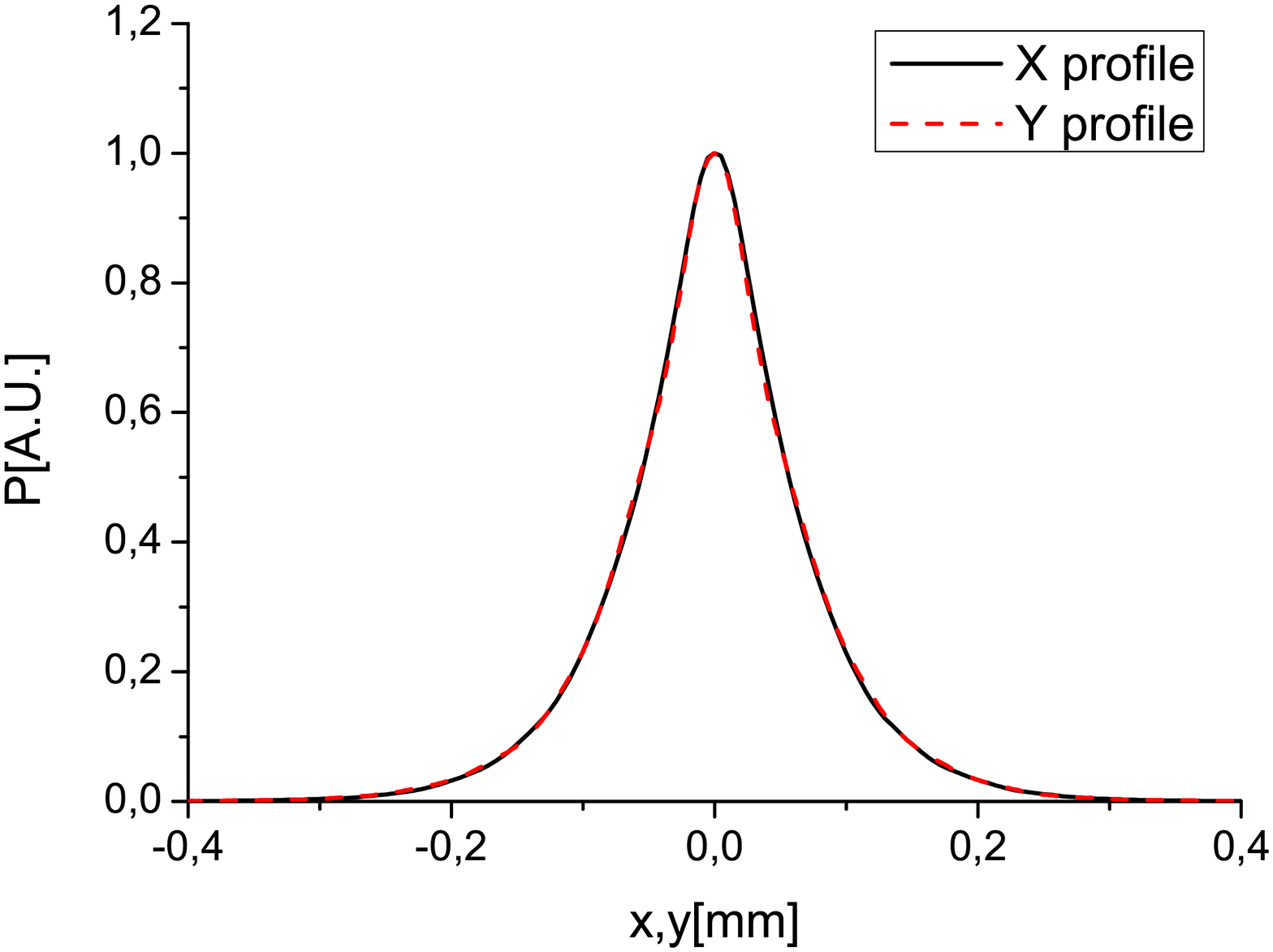}
\caption{(Left plot) Distribution of the radiation pulse energy per
unit surface and (right plot) angular distribution of the radiation
pulse energy at the exit of the output undulator.} \label{spotT}
\end{figure}
Finally, the distribution of the radiation pulse energy per unit
surface and the angular distribution of the radiation pulse energy
at the exit of output undulator are shown in Fig. \ref{spotT}.

\section{\label{sec:potential} Potential for biomolecular imaging with 10 TW-power level x-ray
pulses}

The most interesting novel properties of the source proposed in this
article, which are most important for life science applications, are
the extremely short pulse width (about $10$ fs) and the very high
peak power (about $10$ TW). Imaging of single molecules at atomic
resolution using radiation from the European XFEL facility would
enable a significant advance in structural biology, because it would
provide means to obtain structural information of large
macromolecular assemblies that cannot crystallize, for example
membrane proteins. The imaging method "diffraction before
destruction" \cite{HAJD}-\cite{BERG} requires pulses containing
enough photons to produce measurable diffraction patterns and short
enough to outrun radiation damage. The highest signals are achieved
at the longest wavelength that supports a given resolution, which
should be better than 0.3 nm. These considerations suggest that the
ideal wavelength range for single biomolecule imaging spans between
3 keV and 5 keV \cite{BERG}.

After interacting with a single XFEL pulse the sample is completely
destroyed, so that one molecule can only yield a single measurement.
In order to actually perform the single molecule imaging, several
steps have to be taken. First, a series of single molecules with the
same structure are injected into vacuum and exposed to ultrashort
x-ray pulses. Many diffraction images of the molecule with unknown
orientation are recorded before radiation-induced damage takes
place. This process is repeated until a sufficient number of images
are recorded. Next, the relative orientations of the different
images are determined in order to assemble a 3D diffraction pattern
in the reciprocal space \cite{HULD}-\cite{IKED}. Finally, the 3D
electron density of the molecule is obtained by a phase retrieval
method. The higher the intensity is, the stronger the diffraction
signal, and the higher the resolution for each 2D diffraction
pattern. It can be seen that structural determination of
biomolecules of around 10 nm size require a pulse fluence of about
$10^{22}$ photons/mm$^2$, a resolution of 0.3 nm and a photon energy
around 4 keV. Bio-imaging capabilities can be obtained by reducing
the pulse duration to 10 fs or less, and simultaneously by
increasing the number of photons per pulse to about $10^{14}$. This
yields the required fluence with a 100 nm focus\footnote{We are
assuming that beamline and focusing efficiency are such, that all
$10^{14}$ photons are transmitted into the focus.}.

The key figure for optimizing a photon source for single biomolecule
imaging is the peak power. Ideally, the peak power should be of the
order of 10 TW. In order to motivate this number with an example, we
note that $10^{14}$ photons at 3.5 keV correspond to an energy of
about $60$ mJ which yields, in 10 fs, a peak power of about $6$ TW.
It is worthwhile to mention that 1 TW at 3 keV gives the same signal
per Shannon pixel as 20 TW at 8 keV (assuming a fixed pulse
duration).

\begin{figure}
\begin{center}
\includegraphics[clip,width=0.50\textwidth]{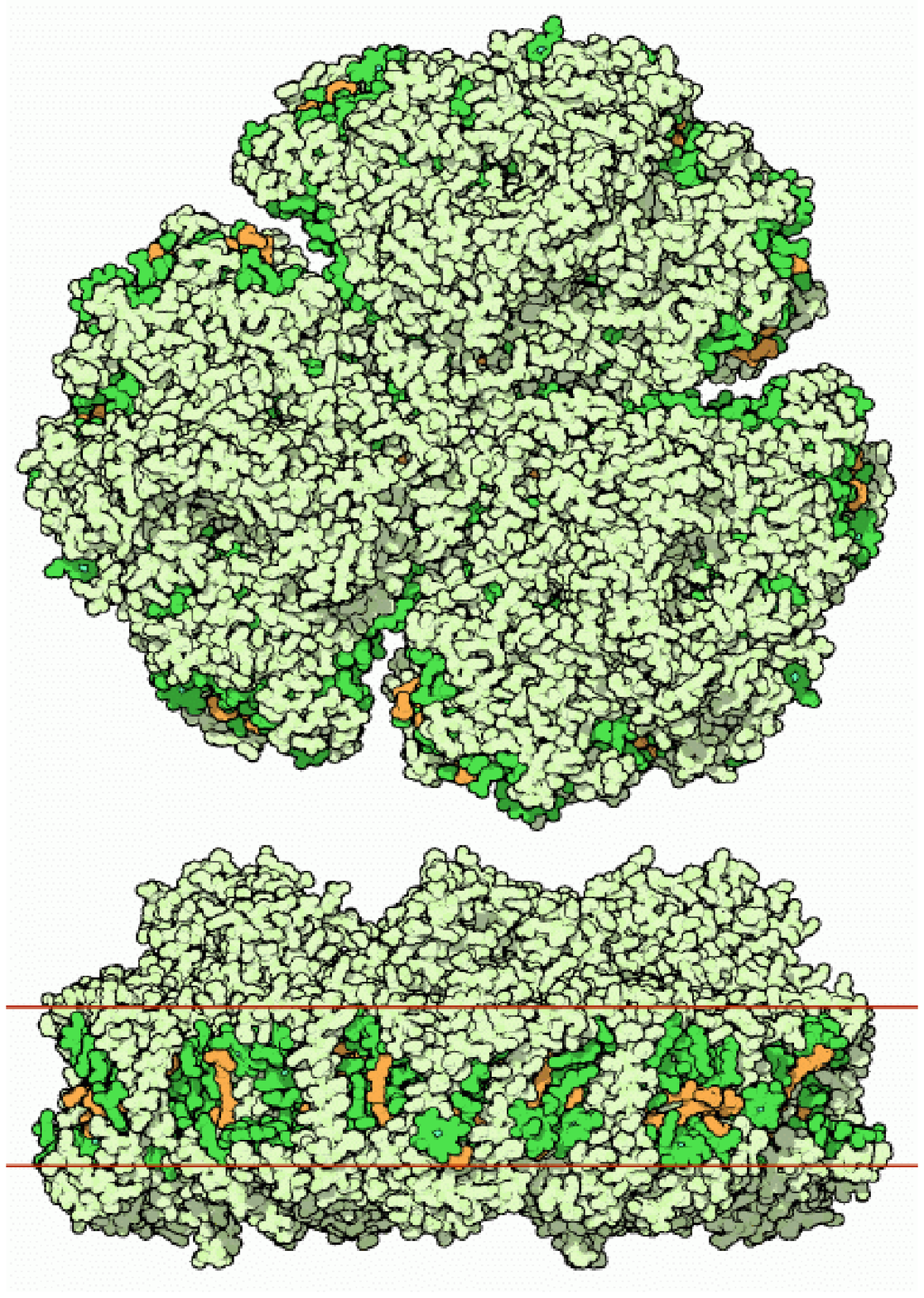}
\end{center}
\caption{Atomic representation of the photosystem-I molecule
\cite{PDBR}} \label{sys1}
\end{figure}
\begin{figure}
\begin{center}
\includegraphics[clip,width=0.50\textwidth]{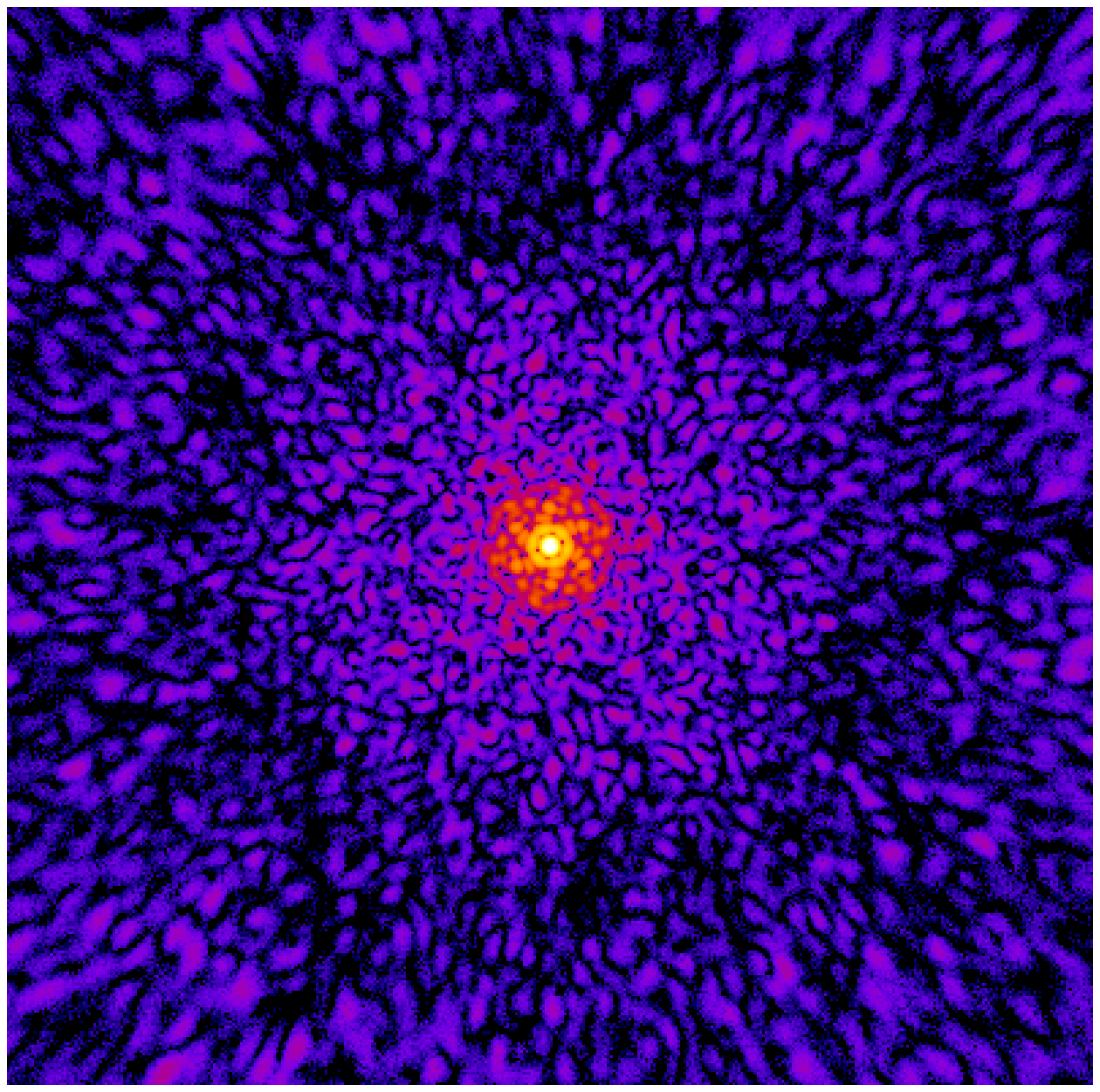}
\end{center}
\caption{Simulated diffraction pattern from a photosystem-I molecule
for a fluence of $10^{22}$ photons/mm$^2$. The simulation\cite{WECK}
was performed for 3.5 keV radiation, neglecting radiation damage.
The detector considered here is $200$ mm by $200$ mm in size, with
$400$ by $400$ pixels, and is located at $100$ mm distance from the
sample.} \label{diffsys}
\end{figure}
\begin{figure}
\begin{center}
\includegraphics[width=0.75\textwidth]{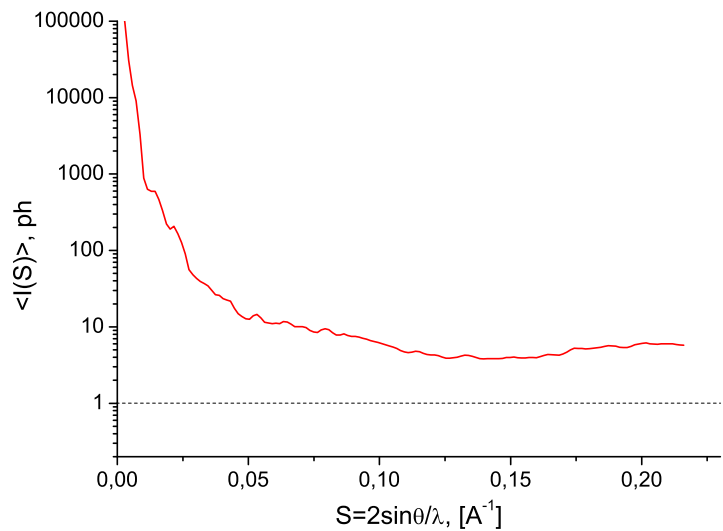}
\end{center}
\caption{Radially averaged scattering intensity as a function of the
modulus of the scattering vector for a photosystem-I molecule
illuminated with $3.5$ keV radiation.  } \label{diff2}
\end{figure}
Simulations confirm that with $10^{14}$ photons in a $10$ fs pulse
at a photon energy around 4 keV and with a 100 nm focus, one can
achieve diffraction to the  desired resolution of $0.3$ nm for a
molecule of about 10 nm size. This fact is exemplified by using the
photosystem-I membrane protein as a case study, Fig. \ref{sys1} .
The simulated diffraction pattern from the photosystem-I for a
fluence of $10^{22}$ photons/mm$^2$ is shown in Fig. \ref{diffsys}.
The simulation was performed neglecting radiation damage. While some
controversy is present in the community concerning the upper limit
to the fluence imposed by radiation damage issue, there are
indications that the diffraction signal still improves up to about
$10^{14}$ photons focused down to $100$ nm \cite{LORE}.

First, the diffraction patterns must be oriented with respect to
each other in 3D Fourier space. The key figure here is the number of
photons per Shannon pixel per single-shot diffraction pattern.  Fig.
\ref{diff2} shows the radially-averaged scattered intensity as a
function of the modulus of the scattering vector. One can see that
most detector pixel values are considerably higher than one photon
count, up to the edge of the detector. Studies have shown that a
signal in the order of one photon per pixel would be sufficient to
correlate diffraction images of identical molecules. A detector
pixel value larger than 1 photon/pixel results in an increase in the
number of classified images, up to the number of hits. For a
molecule of 10 nm size one needs about 100 evenly spread 2D
projections to get a geometrical resolution of 0.3 nm. Thus, at an
average pixel value larger than 1 photon/pixel, a number of hits of
about 100  is required to achieve full 3D information accounting
only  for the photon shot noise.

\section{\label{sec:cons} Conclusions}

The main expectation and the main challenge emerging from
applications of XFEL sources to life sciences, is the determination
of 3D structures of biomolecules and their complexes from
diffraction images of single particles. Only two facilities,
European XFEL \cite{ETDR} and the LCLS-II \cite{LCL2} will have the
possibility to build a beamline suitable for single biomolecule
imaging. In fact, in the next decade, no other XFEL source will have
a driver with high-enough electron beam energy (13-17 GeV) and, at
the same time, long enough undulators (250 m) to enable 10 TW mode
of operation with about 10 fs-long x-ray pulses. In this work we
proposed a cost-effective upgrade program for an undulator source at
the European XFEL, which could result in a beamline optimized for
single biomolecular imaging, with the potential to secure a
world-leading position in the field of life sciences for the
European XFEL.

\section{Acknowledgements}

We are grateful to Massimo Altarelli, Reinhard Brinkmann, Henry
Chapman, Janos Hajdu, Viktor Lamzin, Serguei Molodtsov and Edgar
Weckert for their support and their interest during the compilation
of this work. We thank Y. Ding and Z. Huang for useful discussions
about the emittance spoiler method, Edgar Weckert for providing the
code MOLTRANS to one of us (O.Y.).


\begin{thebibliography}{99}


\bibitem{HAJD} J. Hajdu, Curr. Opin. Struct. Biol. 10, 569 (2000).

\bibitem{NEUT} R. Neutze et al., Nature 406, 752 (2000).

\bibitem{CHAP} K. J. Gaffney and H. N. Chapman, Science 316, 1444
(2007).

\bibitem{SEIB} M. M. Seibert et al., Nature 470 (7332) 78-81 (2011).

\bibitem{BERG} S. Baradaran et al., LCLS-II New Instruments Workshops Report, SLAC-R-993
(2012), see Section 4.3.2. by H. Chapman et al., and Section 4.3.3.
by F. R. N. C. Maia et al.

\bibitem{ETDR} M. Altarelli, et al. (Eds.)
XFEL, The European X-ray Free-Electron Laser, Technical Design
Report, DESY 2006-097, Hamburg (2006).

\bibitem{SELF} J. Feldhaus et al., Optics. Comm. 140, 341 (1997).

\bibitem{SXFE} E. Saldin, E. Schneidmiller,  Yu. Shvyd'ko and M.
Yurkov, NIM A 475 357 (2001).

\bibitem{SOPT} E. Saldin, E. Schneidmiller and M. Yurkov, NIM A 445
178 (2000).


\bibitem{STTF} R. Treusch, W. Brefeld, J. Feldhaus and U Hahn,  Ann. report 2001
"The seeding project for the FEL in TTF phase II" (2001).

\bibitem{SCOM} A. Marinelli et al., Comparison of HGHG and Self Seeded Scheme for the Production of Narrow Bandwidth FEL
Radiation, Proceedings of FEL 2008, MOPPH009, Gyeongju (2008).

\bibitem{OURL} G. Geloni, V. Kocharyan and E.~Saldin, "Scheme for generation of highly monochromatic X-rays from a baseline
XFEL  undulator", DESY 10-033 (2010).


\bibitem{HUAN} Y. Ding, Z. Huang and R. Ruth, Phys.Rev.ST Accel.Beams, vol. 13, p. 060703 (2010).

\bibitem{OURX} G. Geloni, V. Kocharyan and E.~Saldin, "A simple method for controlling the line width of SASE X-ray FELs",
DESY 10-053 (2010).

\bibitem{OURY2} G. Geloni, V. Kocharyan and E.~Saldin, "A Cascade self-seeding scheme with wake monochromator for narrow-bandwidth X-ray FELs", DESY 10-080 (2010).


\bibitem{OURY4} Geloni, G., Kocharyan, V., and Saldin, E., "Cost-effective way to enhance the capabilities of the LCLS baseline", DESY 10-133 (2010).




\bibitem{WU} J. Wu et al., "Staged self-seeding scheme for narrow
bandwidth , ultra-short X-ray harmonic generation free electron
laser at LCLS", proceedings of 2010 FEL conference, Malmo, Sweden,
(2010).


\bibitem{OURY6} Geloni, G., Kocharyan V., and Saldin, E., "Generation of doublet spectral lines at self-seeded X-ray FELs", DESY 10-199
(2010), and Optics Communications,  284, 13, 3348 (2011)


\bibitem{OURY5} Geloni, G.,  Kocharyan, V.,  and Saldin, E., "Production of transform-limited X-ray pulses through
self-seeding at the European X-ray FEL", DESY 11-165 (2011).


\bibitem{OURY5b} Geloni, G., Kocharyan V., and Saldin, E., "A novel Self-seeding scheme for hard X-ray FELs", Journal of Modern
Optics, vol. 58, issue 16, pp. 1391-1403,
DOI:10.1080/09500340.2011.586473 (2011)




\bibitem{WUFEL2} J. Wu et al., Simulation of the Hard X-ray
Self-seeding FEL at LCLS, MOPB09,  FEL 2011 Conference proceedings,
Shanghai, China (2011).

\bibitem{AMAN} J. Amann et al., Nature Photonics, DOI:
10.1038/NPHOTON.2012.180 (2012).

\bibitem{SHVI} R. R. Lindberg and Yu.V. Shvyd'ko, Phys. Rev. ST Accel. Beams 15,
100702 (2012).

\bibitem{FENG3} Y. Feng et al., "System design for self-seeding the
LCLS at soft X-ray energies", Proceedings of the 24th International
FEL Conference, Nara, Japan (2012).

\bibitem{GRAT} S. Serkez, G. Geloni, V. Kocharyan and
E. Saldin, "Grating monochromator for soft X-ray self-seeding the
European XFEL", DESY 13-040, http://arxiv.org/abs/1303.1392 (2013).


\bibitem{ASYM} G. Geloni, V. Kocharyan, E. Saldin, S. Serkez and M.
Tolkiehn, "Wake monochromator in asymmetric and symmetric Bragg and
Laue geometry for self-seeding the European XFEL", DESY 13-013
(2013).


\bibitem{EMM1} P. Emma, K. Bane, M. Cornacchia, Z. Huang, H. Schlarb, G. Stupakov, and D. Walz, PRL 92, 074801-1
(2004).

\bibitem{EMM2} P. Emma, M. Borland and Z. Huang, Proceedings of the FEL Conference
2004, TUBIS01, p.333 (2004).

\bibitem{DING} Y. Ding , et al., PRL 109, 254802-1 (2012).

\bibitem{TAP1} A. Lin and J.M. Dawson, Phys. Rev. Lett. 42 2172
(1986).

\bibitem{TAP2} P. Sprangle, C.M. Tang and W.M. Manheimer, Phys. Rev. Lett. 43 1932
(1979).

\bibitem{TAP3} N.M. Kroll, P. Morton and M.N. Rosenbluth, IEEE J. Quantum Electron., QE-17, 1436
(1981).

\bibitem{TAP4} T.J. Orzechovski et al., Phys. Rev. Lett. 57, 2172
(1986).

\bibitem{FAWL} W. Fawley et al., NIM A 483 p 537 (2002).

\bibitem{CORN} M. Cornacchia et al.,  J. Synchrotron rad. 11,
227-238 (2004).

\bibitem{WANG} X. Wang et al., PRL 103, 154801 (2009).


\bibitem{OURY3} G. Geloni, V. Kocharyan and E.~Saldin, "Scheme for generation of fully coherent, TW power level hard x-ray pulses from baseline undulators at the European XFEL", DESY 10-108 (2010).




\bibitem{WUFEL1} W.M. Fawley et al., Toward TW-level LCLS radiation
pulses, TUOA4, FEL 2011 Conference proceedings, Shanghai, China
(2011).

\bibitem{LAST} Y. Jiao et al. Phys. Rev. ST Accel. Beams 15, 050704
(2012).






\bibitem{S2ER} I. Zagorodnov, "Beam Dynamics Simulations for XFEL",
http://www.desy.de/xfel-beam/s2e (2011), and "Compression scenarios
for the European XFEL",
http://www.desy.de/fel-beam/data/talks/files/Zagorodnov$\_$ACC2012$\_$ready$\_$new.pptx,
(2012).


\bibitem{BIO1} G. Geloni, V. Kocharyan and E.~Saldin, "Conceptual design of an undulator system for a dedicated bio-imaging beamline at the European X-ray FEL",
DESY 12-082, http://arxiv.org/abs/1205.6345 (2012).


\bibitem{BIO3} G. Geloni, V. Kocharyan and E.~Saldin, "Optimization of a dedicated bio-imaging beamline at the European X-ray
FEL", DESY 12-159, http://arxiv.org/abs/1209.5972 (2012).

\bibitem{ELEG} M. Borland, Elegant, http://www.aps.anl.gov/Accelerator$\_$ Systems$\_$Division/Accelerator $\_$Operations$\_$Physics/software.shtml$\sharp$elegant


\bibitem{AUTH} A. Authier, Dynamical Theory of X-Ray Diffraction (Oxford
University, 2001).

\bibitem{GENE} S. Reiche et al., Nucl. Instr. and Meth. A 429, 243 (1999).


\bibitem{HULD} G. Huldt and A. Szoke and J. Hajdu, J. Struct. Biol.,
144, 219 (2003)

\bibitem{BOR1} G. Bortel and G. Faigel J. Struct. Biol. 158, p10
(2007).



\bibitem{FUNG} R. Fung,  V. Shneerson, D. Saldin and A. Ourmazd,
Nature Physics 5, 64 (2009).

\bibitem{LOH1} N. D. Loh et al., Phys. Rev. Lett. 104, 22, 2255051
(2010).



\bibitem{IKED} S. Ikeda and H. Kono, Optics Express 20 p3375 (2012).


\bibitem{PDBR} Protein Data Bank, http://www.rcsb.org/pdb/101/motm.do?momID=22

\bibitem{WECK} Results were obtained using the code MOLTRANS,
written by E. Weckert.

\bibitem{LORE} U. Lorenz,  N. M. Kabachnik,  E. Weckert and I. A.
Vartanyants, Phys. Rev. E" 86, 051911 (2012).



\bibitem{LCL2} The LCLS-II CDR,
https://portal.slac.stanford.edu/sites/lcls$\_$public/lcls
$\_$ii/Published$\_$Documents/CDR$\%$20Index.pdf (2011).

\end{thebibliography}
\end{document}